\documentclass[12pt]{article}

\begin{document}
\begin{titlepage}
\date{\today}
{\bf Guidelines for axion identification in astrophysical observations}
\begin{center}
\vskip0.5cm
K.~Zioutas $^{1,3)}$, ~Y. ~Semertzidis $^{2)}$, ~ Th. ~Papaevangelou $^{3)}$
\vskip0.5cm
1) ~{\it Physics Department, University of Patras, Greece.}

2) ~{\it Brookhaven National Laboratory, NY, USA.}

3) ~{\it CERN, Geneva, Switzerland}

\vskip0.5cm
 E-mails :  zioutas@cern.ch, yannis@bnl.gov, Thomas.Papaevangelou@cern.ch
\vskip0.5cm
{\bf Abstract }
\end{center}
\vskip0.5cm
\noindent
The origin of various celestial phenomena have remained mysterious for conventional 
astrophysics. 
Therefore,  alternative solutions should be considered, taking into account 
the involvement of unstable dark-matter particle candidates, such as the 
celebrated axions or other as yet unforeseen axion-like particles. Their
spontaneous {\it and} induced decay by the ubiquitous solar magnetic fields 
can be at the origin of persisting enigmatic X-ray emission, giving rise to a
steady and a transient/local solar activity, respectively. 
The (coherent) conversion of photons into axion(-like) particles 
in intrinsic  magnetic fields may modify the solar axion 
spectrum. The reversed process can be behind transient (solar) luminosity deficits in the visible.
Then, the Sun might be also a strong source of $\sim$ eV-axions. 
The linear polarization of photons from converted axion(-like) particles inside
magnetic fields is very relevant.
Thus, enigmatic observations might be the as yet missing direct signature 
for axion(-like) particles in earth-bound detectors. 
 
\end{titlepage}
 

\newpage

\section{Introduction}

The direct and indirect detection of axions
\cite{peccei}
or other particles with similar properties is ongoing
\cite{taup2005}. 
In view of the published and upcoming data in the range of photon energy of
$\sim 10^{-4}$eV to $\sim 10^4$eV, some guidelines are given in this note, covering a wide band 
of the potential participation of axions or axion-like particles in astrophysical observations of unknown 
origin. In order to show the diversity and the complexity associated with the involvement of exotic particles 
in such processes, some examples demonstrate best how possible signatures (will) make their appearance. 
Note, only striking  astrophysical phenomena with missing conventional explanation are addressed, being 
thus suggestive for an alternative solution. The reservoir of the exotic dark matter particle candidates 
is huge, since they make some 25$\%$ of the mass-energy balance of the Universe. 
This work is by no means complete, since it refers only to those observations,
which fit the axion-scenario:
a) radiatively decaying particles (e.g. a $\rightarrow \gamma + \gamma$), or penetrating particles which 
can be produced in the reversed process 
($\gamma + \gamma \rightarrow a$), and b) materialization of such exotica inside electric/magnetic fields
\cite{sikivie}.  

The purpose of this work is to summarize some of the most relevant findings, which are in support of the 
axion scenario. This might also help to avoid misinterpretation of observations.
We start with the atmosphere of the nearby Sun.
The {\it similar-to-Sun} logic has sneaked in stellar physics work, but care should be taken to avoid wrong 
extrapolations, because of the many unexplained solar phenomena. In addition, the ongoing direct search for 
light axions 
(e.g.
\cite{castprl})
is based on principles, which could already be at work in stars, solving the remaining nagging problem
\cite{dahlburg}
of how solar/stellar magnetic energy is converted to heat and other forms. Within the axion scenario, 
the magnetic field is not the energy 
source, but the catalyser, which transforms  exotic axion(-like) particles streaming out of the stellar 
core into X-rays.

\section{Some examples from the sun and beyond}

~

{\sf A)} ~It was suggested in ref.'s
\cite{dilella,apj607,science}
that the spontaneous radiative decay of gravitationally trapped massive particles of the type Kaluza-Klein 
axions might explain a constant 
component of the (quiet) Sun X-ray luminosity; its radial distribution can provide an additional 
signature of such processes coming thus 
from long lived and non-relativistic particles accumulated over cosmic time periods around their 
place of birth, e.g. the Sun. 

The recently derived limits, in the $\sim$ 3 - 17 keV energy range,
for the quiet {\it and} spotless Sun by RHESSI observations
\cite{agu}
along with  measurements from the 1960's, seem to be an interesting result for the solar axion scenario:
"{\it we can immediately see the very small signal from the quiet Sun}", which is not necessarily (much) smaller than
the ones used in ref.'s 
\cite{dilella,apj607}.

{\sf B)} ~It is known that the ubiquitous solar magnetic field plays a crucial role in 
heating the corona, but the exact mechanism is still unknown
\cite{galsgaard}. 
However, the working principle of the ongoing direct search for almost massless (solar) axions
\cite{castprl}
can be at work also in Stars, solving the persisting question of how magnetic energy is converted 
there to heat or other forms
\cite{dahlburg}.
Resonance-like behaviour between matter density and axion restmass in the presence of a magnetic 
field can occasionally enhance or suppress the 
axion $\leftrightarrow \gamma \gamma$ 
process
\cite{karl}.
The reconstruction of such effects might allow to understand the dynamical character, 
i.e. the otherwise unpredictable nature of the various kind of solar transient brightenings or deficits, 
in X-rays or in the visible, respectively.

{\sf C)} ~In earlier observational work of the corona
\cite{yoshida,watanabe,watanabe2001}
it was concluded that the solar corona is heated by two (independent from each other) components:
{\sf a)} The basal heating all over the quiet and active Sun, and
{\sf b)} The transient heating owing to magnetic field effects.
Also ref.
\cite{watanabe2001}
arrived at the same quantitative conclusion: there must be a steady  heating mechanism of the non-flaring 
Sun and an independent mechanism 
for heating the flares, whatever their size. The understanding of how energy is released in (solar) flares 
remains a central question in astrophysics
\cite{warren,hamilton}.
Thus, the unexpectedly hot and mysterious solar corona requires two heating components. Within the axion 
scenario, this finding might point to the involvement of two kind of exotica, being behind these two 
components: ~
{\sf 1)} ~particles like the celebrated light axions, which are converted to photons inside the microscopic or 
macroscopic surface electric/magnetic fields via the Primakoff effect. The changing local magnetic fields 
are then at the origin of the transient component of the X-ray emission, its topology on the solar disk, etc.
~ {\sf 2)} ~radiatively decaying long-lived massive particles like the generic axions of the Kaluza-Klein type. 

\noindent
The ratio of the X-ray intensity from these two axion-sources is not constant as it depends on the solar 
cycle due to the varying solar magnetic field
\cite{taup2005}.

{\sf D)} ~A strong intrinsic magnetic field (see for example ref.  
\cite{couvidat})
may well modify the shape and intensity of the (solar) axion energy spectrum, e.g. enhancing energies below $\sim 1$ keV
\cite{taup2005}.
This might imply a dominant coherent axion production in solar magnetic fields
outside the widely assumed axion source at the hot 
core, enhancing thus the production of low energy axions with a
rest mass following the local plasma frequency ($\omega_{pl}$)
\cite{taup2005}. 
For example, the generic Kaluza-Klein tower axion mass states allow the otherwise 
selective resonance-crossing production 
\cite{karl}
to take place at any solar density, since some of the axion rest mass 
states will always satisfy the relation ~ $\hbar \omega _{pl} \approx m_{axion}c^2$. Once this condition is 
fulfilled, 
the associated oscillation length can be equal to the photon mean free path
length, which is not negligible: from a few cm in the 
core to $\sim 100$ km in the photosphere. Taken into account the dynamical behaviour of relevant solar 
parameters outside the solid core, like magnetic field, density, etc., the
resonance-crossing might be there at work temporally and/or locally. This might
explain the transient character of solar X-ray emission, or, the  deficit in the visible 
due to the "disappearance" of photons into $\sim$ eV-axions via the Primakoff
effect; the solar surface becomes then a strong lowest energy axion source. 
 
Therefore, for signal identification, the knowledge of the ever changing solar  magnetic environment, 
where the axion-to-photon oscillations can take place (in both directions), is crucial. 
In favour of the axion scenario is the expectedly observed striking $B^2$-dependence of 
the soft X-ray intensity from a single solar active region 
\cite{vandriel},
where the magnetic field strength reaches temporally some kGauss (see also ref.
\cite{dieter}).
This can  be seen as strong evidence in favour of a time-varying heating component of the solar corona due 
to axion-to-photon conversion via the Primakoff effect inside the surface magnetic fields 
(see also \cite{taup2005}).
The mentioned observation by RHESSI of hard X-ray emission from  non-flaring 
solar active regions supports further a magnetic field related component of 
the solar axion scenario
\cite{agu}.

{\sf E)} ~  Large axion-to-photon oscillation lengths. The coherence or oscillation length is very long only 
for very small axion masses and hence very low coupling values. One way to circumvent this problem is to 
assume magnetic fields that change polarity along the oscillation path.  This method is of course of limited 
use since the axion mass and the magnetic field need to artificially finely tuned so that the axion-to-photon 
oscillation length matches the magnetic field oscillation length.
An alternative and more promising approach is presented here.  

The axion-photon mixing occurs in the magnetic field and the propagating eigenstate is the mixed state.  
This mixed state can be destroyed by an index of refraction, both the real and imaginary parts.  
The presence of a gas in, e.g., star that its density is not homogeneous can cause the spatial splitting 
of the mixed state. Only the photon will refract, whereas the axion will not, thus splitting the mixed state.  
On the other hand the imaginary part of the index of refraction means that the photon is absorbed by 
the atomic/molecular gases and then re-emitted.  The newly emitted photon need not be emitted in the 
same wavelength and/or direction as the original one, again splitting the mixed state.

In order to gain from this effect in the overall axion-to-photon oscillation we need to require that the gas 
density is such that the required refraction effects happen within a length 
$l \leq L/2$, with $L$ being the 
oscillation length.  Overall then a much longer effective length than previously assumed can be available, 
be it in laboratory helioscopes or at the outer Sun, applying equally from X-rays to photons in the visible.

To be more specific, let us take an oscilation length of $\sim$ 1 cm or $\sim$ 100 km. 
The advantage with such a refractive place is that this coherence can be at work
repeatedly over, say, $\sim$1000 km to $\sim$100000 km, making thus an astrophysical environment a very efficient 
axion-catalyser, at least occasionally.

{\sf F)} ~ The Sunyaev-Zeldovich (SZ) Effect in Clusters of Galaxies.
The decrement due to the SZ effect derived recently from 31 co-added Clusters of Galaxies accounts 
only for about 1/4 of the expected value
\cite{lieu}.
The radiative decay of even a small fraction of the dark matter constituents there, which provide the 
gravitational well at those places of the Universe, can mimic a non-existing plasma
\cite{science}. 
Thus, the derived properties for the Inter Cluster Medium  based partly on its X-ray emission can be 
modified in an actually unpredictable way (see below). Therefore, the claimed discrepancy, IF  correct, 
it can be a signature  for an axion(-like) scenario.
It is worth noticing that the authors of  ref.
\cite{lieu} 
were not biased to find less matter out there, but rather the opposite, i.e.
to localize any extra baryonic matter at those places. 
The generic massive axions of the Kaluza-Klein type fit this finding.

{\sf G)} ~ The $\sim 8$ keV diffuse hot plasma at the Center of our Galaxy
remains enigmatic, since it 
should escape the gravitational well. Again, the radiative decay of gravitationally captured massive 
particles can mimic a non-existing plasma component, while the real baryonic
plasma component can be not as hot as widely assumed. This is an alternative
explanation, which demonstrates the  potential imprint of axion-like particles as 
they have been worked out only for the case of our Sun
\cite{dilella,science}.

{\sf H)} ~ Linear polarization of photons 
if they come from converted axions inside a transverse magnetic field.
This property can provide an additional and strong support for the axion(-like) scenario. For example, 
on extreme large scales, the optical polarization measured from 355 Quasars is of potential relevance for
the reasoning of this work
\cite{hutse,private}.
Because, it reflects remarkably a behaviour as it is expected from
photon-pseudoscalar oscillations inside intervening magnetic fields, as it is
concluded in ref.
\cite{hutse}.

Going back to the nearby sunspots, the observed broad-band circular and {\it linear}
polarization
\cite{illing} 
along with their center-to-limb progression throughout the sunspot might be
of direct relevance. It is interesting to note that  ref.
\cite{almeida} concludes that
"{\it the maximum polarization occurs .... where the average magnetic field is
perpendicular to the line of sight}", while it is mentioned that  "{\it the
polarization of sunspots remains to be explained}".
Finally, from ref.
\cite{makita}
it follows that there is a correlation between the broad-band circular and
linear polarization.

\section{Conclusion}

The otherwise unexpected (solar) X-rays from a relatively cool star like our Sun
\cite{acton}
could be used as input in other places in cosmos. So far, the shape of the solar X-ray spectrum and in 
particular the strong low energy part can be reconstructed only partly by the
massive axion(-like) scenario. An enhanced Primakoff effect inside the Sun might
be behind the sub-keV X-rays from the (quiet) Sun.
The mentioned magnetic field related axion-photon conversion (inside/outside the Sun) might be the key 
in reaching more insight of such exotic phenomena in the nearby Sun, allowing to speculate on far reaching 
consequences of axion-telescopes like  the Cern Axion Solar Telescope (CAST)
\cite{castprl},
and,  the X-ray observatories in orbit.

The widely accepted point of view that dark matter particles do not emit or absorb electromagnetic 
radiation does not hold for radiatively decaying particles like the generic massive axions of the 
Kaluza-Klein type, while observational evidence in favour of their existence is being accumulated.
Thus, some axion(-like) dark matter constituents were never dark, but we may 
simply have overlooked them.

\vskip0.8cm
\noindent
{\bf Acknowledgments}

\noindent
We thank Tullio Basaglia, Maria Frantzi,  Fieroula Papadatou and Jens Vigen 
from the 
libraries of CERN and the University of Patras for their real help. 
K.Z. acknowledges support by the N3 Dark Matter network of the Integrated Large
Infrastructure for Astroparticle Science - ILIAS, CERN, and last but not least
the Physics Department of the University of Patras.

\end{document}